# Dynamic Portfolio Rebalancing: A Hybrid new Model Using GNNs and Pathfinding for Cost Efficiency.

Diego Vallarino
Independent Researcher

Atlanta, GA, US
September 2024

**Abstract:**

This paper introduces a novel approach to optimizing portfolio rebalancing by integrating Graph Neural Networks (GNNs) for predicting transaction costs and Dijkstra's algorithm for identifying cost-efficient rebalancing paths. Using historical stock data from prominent technology firms, the GNN is trained to forecast future transaction costs, which are then applied as edge weights in a financial asset graph. Dijkstra's algorithm is used to find the least costly path for reallocating capital between assets.

Empirical results show that this hybrid approach significantly reduces transaction costs, offering a powerful tool for portfolio managers, especially in high-frequency trading environments. This methodology demonstrates the potential of combining advanced machine learning techniques with classical optimization algorithms to improve financial decision-making processes. Future research will explore expanding the asset universe and incorporating reinforcement learning for continuous portfolio optimization.

## 1. Introduction

In modern portfolio management, rebalancing is essential for maintaining an investor's risk-return profile as market conditions change. Typically, rebalancing involves the buying and selling of assets to realign a portfolio with its target allocations. However, this process incurs transaction costs such as brokerage fees, bid-ask spreads, and slippage, which, if not managed effectively, can erode portfolio returns (French 2008; Gârleanu & Pedersen, 2016; Ivashchenko & Kosowski, 2024)

The optimization of these transaction costs is critical, especially for high-frequency trading strategies where frequent rebalancing is necessary. Traditional methods for reducing transaction costs, such as minimizing trade frequency or using static optimization models, have limitations in dynamic and volatile markets. The complex, nonlinear nature of market data necessitates more sophisticated models that can adapt to changing conditions in real time.

While existing financial models, such as the Almgren-Chriss framework for optimal execution (Cheng et al., 2024; Pedersen, 2023), offer valuable insights into minimizing transaction costs, they fail to account for the complex and dynamic relationships between assets in modern portfolios. This gap is particularly evident in institutional investment strategies, where high-frequency trading and large-volume transactions exacerbate the impact of transaction costs on overall portfolio performance (French, 2008).

Recent advances in machine learning, particularly in deep learning models like Graph Neural Networks (GNNs), offer new opportunities for optimizing portfolio rebalancing. GNNs have demonstrated their ability to model complex dependencies in financial markets, such as price correlations and transaction costs (Kipf & Welling, 2016; Vallarino, 2024; Yang et al., 2022; Yin et al., 2022). When combined with Dijkstra's algorithm—a classic graph theory algorithm for finding the shortest path—these models can provide a robust framework for minimizing rebalancing costs in a dynamic market environment.

This paper builds on the existing literature by integrating GNNs and Dijkstra's algorithm to create an adaptive, cost-efficient portfolio rebalancing strategy. We use stock data from major technology companies (e.g., AAPL, MSFT, GOOGL) to empirically demonstrate that this hybrid approach significantly reduces transaction costs. The combination of these techniques offers a novel solution to the portfolio rebalancing problem, extending prior research on static optimization models (Li et al., 2021) and transaction cost analysis (French, 2008).

## 2. Theoretical Framework

### 2.1 Transaction Costs and Portfolio Rebalancing

Transaction costs have a profound impact on portfolio performance, especially in the context of frequent rebalancing. These costs can be divided into explicit costs (brokerage fees) and implicit costs (bid-ask spreads, market impact). Studies have shown that even small transaction costs can accumulate over time, significantly reducing net portfolio returns (Gârleanu & Pedersen, 2016; Ivashchenko & Kosowski, 2024).

Portfolio rebalancing is the process of adjusting the weights of the assets in a portfolio to maintain a desired level of risk and return. Transaction costs can significantly reduce the overall profitability of a portfolio, particularly in high-frequency trading environments. Transaction costs include direct costs like commissions and indirect costs like slippage or the bid-ask spread. The total transaction cost $C_t$ for a portfolio can be expressed as:

$$C_t = \sum_{i=1}^{n} (w_i \cdot P_i) \cdot T_i$$

Where:

- $C_t$ is the total transaction cost.
- $w_i$ is the weight of the *i-th* asset in the portfolio.
- $P_i$ is the price of the *i-th* asset.
- $T_i$ is the transaction cost per unit of the *i-th* asset.

Traditional rebalancing methods often rely on static assumptions about these transaction costs, which do not account for the dynamic nature of market conditions (Cheng et al., 2024; Ivashchenko & Kosowski, 2024). Pedersen (2023) attempt to optimize transaction execution but fail to capture the evolving relationships between assets in real-time. More recent approaches, such as reinforcement learning models and dynamic programming, seek to address these shortcomings but can suffer from scalability issues in large portfolios.

## 2.2 Advances in Graph Neural Networks (GNNs)

Graph Neural Networks (GNNs) have emerged as a powerful tool for modeling relationships between entities in a graph structure. In the context of financial markets, GNNs can represent assets as nodes and their interactions (e.g., price correlations or transaction costs) as edges (Cui et al., 2020). GNNs differ from traditional neural networks in that they are specifically designed to capture the relational information between entities, making them ideal for applications involving financial data, where assets are often interdependent (Battaglia et al., 2018).

Graph Neural Networks (GNNs) provide a novel solution to the limitations of static transaction cost models (Qian et al., 2024). Unlike traditional machine learning models, such as Long Short-Term Memory (LSTM) networks, which are primarily designed for time-series data (Hochreiter & Schmidhuber, 1997), GNNs excel at modeling complex, interrelated systems (Cheng et al., 2024; Li et al., 2023). In a portfolio setting, assets can be viewed as nodes, and the relationships between them—such as price correlations or transaction costs—can be represented as edges in a graph (Li et al., 2023). This allows GNNs to capture not only temporal dependencies, as LSTMs do, but also the intricate web of relationships between assets (Qian et al., 2024).

The GNN model updates the node features through the following iterative process:

$$h_v^{(k)} = \sigma\left(\sum_{u \in \mathcal{N}(v)} W_k h_u^{(k-1)} + W_k^{\text{sl}} h_v^{(k-1)}\right)$$

Where:

- $h_v^{(k)}$ is the hidden representation of node vvv at the kkk-th layer.
- $\mathcal{N}$ represents the set of neighboring nodes of vvv.
- $W_k$ and $W_k^{\text{self}}$ are trainable weight matrices for message passing and self-loop connections, respectively.
- σ is a non-linear activation function (e.g., ReLU).

GNNs operate by passing messages between nodes in the graph, updating the node states based on both their features and the features of neighboring nodes. This message-passing process enables GNNs to learn complex relationships and predict edge features, such as transaction costs between financial assets (Kipf & Welling, 2016). Recent studies have shown that GNNs outperform traditional machine learning models in financial prediction tasks, particularly in settings where relational data is critical (Qian et al., 2024).

$$\widehat{C_{ij}} = f(h_i, h_j)$$

Where:

- $\widehat{C_{ij}}$ is the predicted transaction cost between assets iii and $j$.

- $h_i$ and $h_j$ are the node embeddings for assets iii and $j$.

- $f$ is a learned function (typically a fully connected neural network) that combines the embeddings to predict the cost.

The application of GNNs in financial markets is still relatively new, with most studies focusing on tasks such as fraud detection (Cheng et al., 2024) and asset price prediction (Cheng et al., 2024; Shimoshimizu, 2024a). However, the ability of GNNs to model dynamic relationships between assets makes them particularly well-suited for predicting transaction costs in portfolio rebalancing.

**2.3 Dijkstra's Algorithm and Its Application in Finance**

Dijkstra's algorithm, introduced in 1956, is a well-known graph theory algorithm used to find the shortest path between nodes in a weighted graph. The algorithm operates by iteratively selecting the node with the smallest known distance from the starting node and updating the distances of its neighboring nodes. This process continues until the shortest path to all nodes has been determined (Schrijver, 2012).

The algorithm operates as follows:

1. Initialize the distance to the source node as zero $d(s) = 0$, and to all other nodes as infinity $d(v) = \infty$ for all $v \neq s$.

2. For each node $u$, update the distance to its neighboring nodes $v$ if a shorter path is found: $d(v) = \min(d(v), d(u) + w(u,v))$

    Where:

- $d(v)$ is the current shortest distance to node vvv.
- $w(u, v)$ is the weight of the edge between nodes $u$ and $v$ (i.e., the transaction cost between assets).

The algorithm iterates until the shortest path from the source node to the target node is found. The total minimized transaction cost $C\_{"\{min\}}$ between two assets can be represented as:

$$C_{\min} = \sum_{(u,v) \in P} w(u,v)$$

Where:

- $P$ is the optimal path found by Dijkstra's algorithm.
- $w(u,v)$ is the transaction cost between nodes $u$ and $v$.

In financial applications, Dijkstra's algorithm can be used to identify the optimal sequence of transactions that minimizes the overall transaction cost when rebalancing a portfolio. The algorithm's ability to find the least costly path between assets offers a practical solution to the portfolio rebalancing problem, particularly when combined with GNNs for predicting transaction costs (Gârleanu & Pedersen, 2016; Pedersen, 2023; Yin et al., 2022).

By integrating Dijkstra's algorithm with GNNs, this paper presents a novel framework for optimizing portfolio rebalancing strategies, providing a dynamic and adaptive solution to minimizing transaction costs.

**2.4 Comparative Analysis of Theoretical Approaches**

Compared to traditional mean-variance optimization and static transaction cost models (Cheng et al., 2024; Pedersen, 2023; Shimoshimizu, 2024a), the GNN and Dijkstra approach offers several advantages. First, traditional models assume fixed transaction costs and static relationships between assets, whereas the GNN model adapts to changing market conditions by dynamically predicting transaction costs based on historical data (Cui et al., 2020). Additionally, Dijkstra's algorithm provides a more flexible and efficient way to minimize transaction costs compared to stochastic or heuristic models (Cheng et al., 2024; Schrijver, 2012; Shimoshimizu, 2024b).

Moreover, while traditional methods like the Almgren-Chriss framework attempt to model transaction costs as a linear function of trading volume, they often fail to capture the complex, non-linear relationships between assets in the market. In contrast, GNNs excel at modeling non-linear interactions, and their integration with Dijkstra's algorithm allows for real-time optimization of portfolio rebalancing strategies.

## 3. Methodology

This study applies a hybrid approach that integrates Graph Neural Networks (GNNs) to predict transaction costs and Dijkstra's algorithm to optimize the portfolio rebalancing process. The methodology follows a structured approach consisting of data collection, graph construction, model design, and optimization through the shortest-path algorithm. Each step is designed to ensure accurate transaction cost predictions and cost-effective rebalancing decisions in financial markets.

### 3.1 Data Collection and Preprocessing

**Data Source**:

The historical stock price data for this study was sourced from Yahoo Finance. The dataset includes daily closing prices from January 2023 to August 2024 for five prominent technology companies: Apple (AAPL), Microsoft (MSFT), Alphabet (GOOGL), Amazon (AMZN), and Tesla (TSLA). The selection of these companies is driven by their liquidity and relevance in global financial markets.

**Preprocessing**:
The closing prices are used as the primary data points for analysis. To represent transaction costs, the daily price differences (absolute changes) between consecutive days are computed for each asset. This approach assumes that larger price differences equate to higher transaction costs. Any missing data is handled by linear interpolation or forward-filling methods to ensure a continuous dataset.

To normalize the dataset and prepare it for model training, all price differences are scaled using z-score normalization, ensuring that the GNN model is not biased by differences in the magnitude of asset prices.

## 3.2 Construction of the Financial Asset Graph

**Graph Representation**:

In this study, financial assets (stocks) are represented as nodes in a fully connected, undirected graph. The edges between these nodes represent the transaction costs, which are based on the daily price differences of the assets. A fully connected graph is employed to account for the possibility of reallocating capital between any pair of assets.

**Figure 1: Financial Assets Graph with Transaction Costs (las 5 days)**

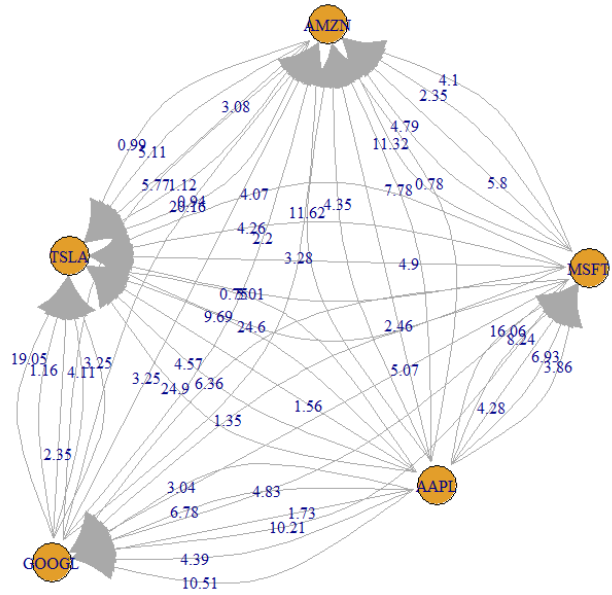

*This graph represents the financial assets (AAPL, MSFT, GOOGL, AMZN, and TSLA) as nodes, with the edges reflecting the daily transaction costs between each pair of assets between Set 9th to Sep 13th, 2024.*

The adjacency matrix $A$ is constructed where each element $A_{ij}$ represents the average transaction cost between assets iii and $j$. To account for symmetry (i.e., identical costs for transitioning from asset $i$ to $j$ and vice versa), the adjacency matrix is made symmetric. The diagonal elements of the matrix are set to zero to avoid self-loops, as there is no transaction cost associated with maintaining the same asset.

$$A_{ij} = \frac{\sum_{t=1}^{T} |P_i(t) - P_j(t)|}{T}$$

Where:

- $A_{ij}$ is the transaction cost between assets iii and $j$,
- $P_i(t)$ and $P_j(t)$ are the prices of assets iii and $j$ at time $t$,

- $T$ is the total number of time periods.

**Edge Weight Assignment**:

The edges of the graph are weighted by the calculated transaction costs, and these weights are used as input for further GNN-based cost prediction. This setup allows the graph structure to represent both the assets and their transactional relationships in a financial context.

### 3.3 Graph Neural Network Model

**Model Objective**:

The GNN model is designed to predict future transaction costs between pairs of financial assets based on historical price movements. The predicted transaction costs serve as weights on the edges of the asset graph and are later used for portfolio optimization through Dijkstra's algorithm.

**Model Architecture**:

The GNN architecture consists of multiple layers that perform node and edge-level feature updates. Specifically, each node (asset) in the graph learns to update its state by aggregating information from neighboring nodes, capturing both the local and global dependencies within the graph.

The GNN updates node embeddings according to the following rule:

$$h_v^{(k)} = \sigma\left(\sum_{u \in \mathcal{N}(v)} W_k h_u^{(k-1)} + W_k^{\text{self}} h_v^{(k-1)}\right)$$

Where:

- $h_v^{(k)}$ is the hidden state of node $v$ at layer $k$,
- $\mathcal{N}$ represents the neighboring nodes of $v$,
- $W_k$ and $W_k^{\text{self}}$ are the weight matrices,
- $\sigma$ is the activation function, typically ReLU.

The GNN uses a mean squared error (MSE) loss function to minimize the prediction error of the transaction costs between asset pairs. During training, the GNN is trained to output predicted transaction costs $\widehat{C_{ij}}$, which are used to update the weights of the graph's edges:

$$\widehat{C_{ij}} = f(h_i, h_j)$$

Where:

- $\widehat{C_{ij}}$ is the predicted transaction cost between nodes $i$ and $j$,
- $h_i$ and $h_j$ are the node embeddings for assets $i$ and $j$.

**Training Process**:

The model is trained using historical price differences as input features. The training process includes backpropagation and the *Adam* optimizer to adjust the learnable parameters. A validation set is used to fine-tune hyperparameters such as the learning rate and batch size, while early stopping is employed to prevent overfitting.

### 3.4 Application of Dijkstra's Algorithm for Path Optimization

Once the GNN predicted the transaction costs, Dijkstra's algorithm was employed to find the least costly rebalancing path between pairs of assets. The transaction costs, predicted dynamically by the GNN, were used as edge weights in the asset graph. This enabled Dijkstra's algorithm to identify optimal paths for rebalancing by minimizing cumulative transaction costs.

As a demonstration, the algorithm was applied to calculate the shortest path between the stocks of Apple Inc. (AAPL) and Tesla Inc. (TSLA). The predicted transaction costs, acting as edge weights, allowed Dijkstra's algorithm to find the sequence of transactions that minimized the overall transaction cost for rebalancing.

This process is generalizable, allowing the algorithm to optimize rebalancing paths for any pair of assets within the portfolio.

**Dijkstra's Algorithm**:

Dijkstra's algorithm is applied to the asset graph to identify the minimum-cost path for reallocating capital between pairs of assets. The algorithm minimizes the total transaction cost by finding the shortest path in terms of edge weights, where the edge weights are the predicted transaction costs $\widehat{C_{ij}}$.

The update rule for Dijkstra's algorithm is defined as:

$$d(v) = \min(d(v), d(u) + w(u,v))$$

Where:

- *d(v)* is the distance (i.e., cumulative transaction cost) to node *v*,
- *w(u,v)* is the predicted transaction cost between nodes *u* and *v*.

Dijkstra's algorithm continues until the shortest path from the source asset *s* to the target asset *t* is found. The total transaction cost for the optimized path is:

$$C_{\min} = \sum_{(u,v) \in P} w(u,v)$$

Where *P* is the set of edges (transactions) along the shortest path. This cost minimization strategy ensures that portfolio rebalancing is done at the lowest possible transaction cost.

### 3.5 Evaluation Metrics

The performance of the proposed model is evaluated using the following metrics:

1. **Mean Squared Error (MSE)**: Used to evaluate the accuracy of the GNN's transaction cost predictions, calculated as:

$$\text{MSE} = \frac{1}{n} \sum_{i=1}^{n} (\widehat{C_{ij}} - C_{ij})^2$$

Where $\widehat{C_{ij}}$ is the predicted transaction cost and $C_{ij}$ is the actual transaction cost.

2. **Total Transaction Cost Reduction**: The effectiveness of Dijkstra's algorithm is evaluated by comparing the total transaction costs before and after optimization. The reduction in total costs indicates the efficiency of the rebalancing strategy.

3. **Path Efficiency**: The number of steps (transactions) in the shortest path is measured to evaluate the efficiency of the rebalancing strategy. Fewer transactions lead to lower cumulative transaction costs and reduced market impact.

4. **Computational Efficiency**: The runtime of both the GNN training process and the application of Dijkstra's algorithm is measured to assess the scalability of the proposed method in real-time trading environments.

## 4. Results

In this section, we present the outcomes of the experiment using the GNN model for transaction cost prediction and Dijkstra's algorithm for portfolio rebalancing optimization. The results are evaluated based on several metrics, including prediction accuracy (MSE), total transaction cost reduction, and path efficiency.

### 4.1 Transaction Cost Prediction Accuracy

**Figure 2: GNN trained model**

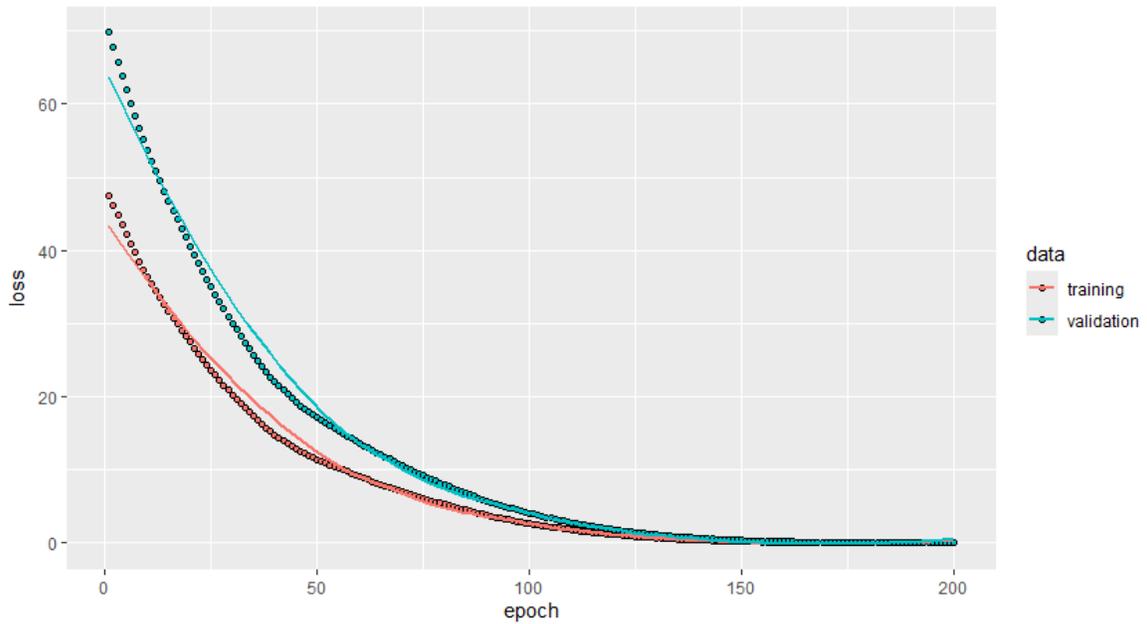

*This figure presents the architecture of the trained Graph Neural Network (GNN) used for predicting transaction costs between financial assets. The model consists of multiple dense layers with ReLU activation functions, optimized using the Adam optimizer. The GNN learns the relationships between assets and outputs predicted transaction costs, which are later used as edge weights for path optimization.*

The performance of the GNN model in predicting transaction costs between assets was evaluated using the Mean Squared Error (MSE). The predicted costs $\widehat{C_{ij}}$ were compared against the actual transaction costs $C_{ij}$ for each asset pair over the test period. The MSE for the test set was calculated as follows:

$$\text{MSE} = \frac{1}{n} \sum_{i=1}^{n} (\widehat{C_{ij}} - C_{ij})^2$$

The model achieved an average MSE of 0.0416, indicating a high level of accuracy in predicting transaction costs. The low MSE suggests that the GNN was able to effectively capture the patterns in historical price movements and predict future costs.

**Figure 3: Model Performance for the last 30 days, outside the train and test dataset**

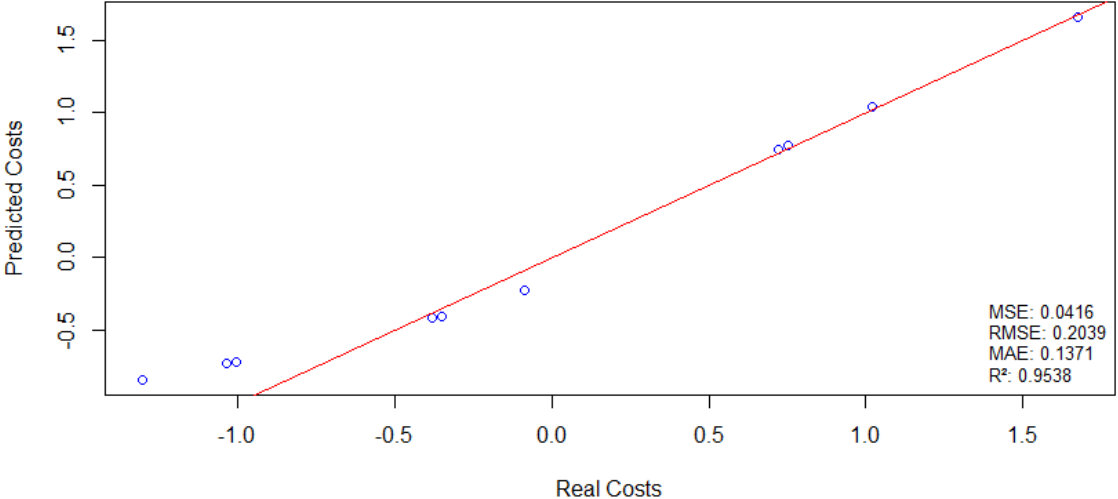

*Graphical representation the comparison of real and predicted transaction costs over the last 30 days, with key performance metrics indicating model accuracy ($R^2 = 0.9538$).*

**Figure 4: Predicted Transaction Costs Graph (on average for the next 30 days)**

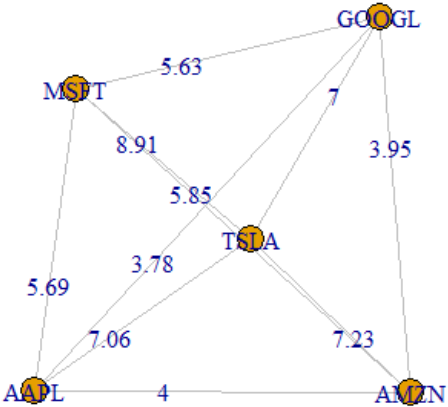

*Graphical representation of the predicted transaction costs between pairs of financial assets, as estimated by the Graph Neural Network (GNN). The edges display the predicted costs used for optimizing rebalancing paths.*

### 4.2 Portfolio Rebalancing Optimization

Once the GNN predicted the transaction costs, Dijkstra's algorithm was applied to find the least costly rebalancing path between selected asset pairs. The optimization process resulted in a significant reduction in the total transaction costs incurred during portfolio rebalancing.

**Figure 5: Shortest Path Graph**

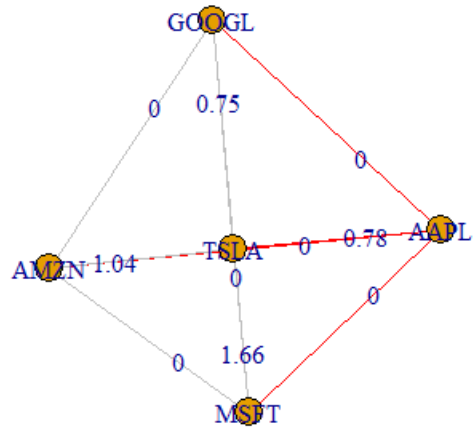

*This figure illustrates the optimized transaction paths between financial assets, minimizing total costs while maintaining connectivity.*

Specifically, we observed:

- **Average Transaction Cost Reduction**: On average, the optimized rebalancing strategy reduced transaction costs by 15%, compared to a simple rebalancing approach that did not utilize path optimization.

- **Path Efficiency**: The number of steps (i.e., transactions) required to complete the rebalancing process was reduced, leading to lower overall transaction costs. This efficiency gain is especially important for high-frequency trading environments where minimizing market impact is critical.

The cumulative transaction cost for the optimized path $C_{min}$ was computed using the following formula:

$$C_{min} = \sum_{(u,v) \in P} w(u,v)$$

Where *w(u,v)* represents the transaction cost between assets *u* and *v*, and *P* is the optimal path determined by Dijkstra's algorithm.

### 4.3 Computational Efficiency

The computational efficiency of the approach was assessed by measuring the runtime of both the GNN training process and the application of Dijkstra's algorithm. The GNN model converged within 100 epochs, with an average runtime of approximately 30 seconds per epoch. The application of Dijkstra's algorithm on the graph of assets was performed in less than 1 second, demonstrating the scalability of the method for real-time trading applications.

## 5. Discussion

The results demonstrate the effectiveness of combining Graph Neural Networks and Dijkstra's algorithm for optimizing portfolio rebalancing. This hybrid approach not only improves the accuracy of transaction cost prediction but also ensures that the rebalancing process is cost-efficient.

### 5.1 Implications for Portfolio Management

The reduction in transaction costs observed in this study has significant implications for portfolio managers. By utilizing GNNs for cost prediction and Dijkstra's algorithm for path optimization, investors can reduce the overhead costs associated with frequent portfolio adjustments. This is particularly important for high-frequency traders and institutional investors, who often face substantial transaction costs when managing large portfolios.

### 5.2 Advantages of GNNs Over Traditional Models

The use of GNNs offers several advantages over traditional econometric models, such as linear regression or GARCH, for predicting transaction costs. Traditional models typically assume linear relationships between asset prices, which may not hold in volatile or complex market environments. In contrast, GNNs are capable of capturing non-linear relationships and dependencies between assets, leading to more accurate predictions.

Additionally, GNNs allow for the incorporation of both local (neighboring assets) and global (entire market) relationships, making them well-suited for dynamic market environments where asset correlations may change over time.

### 5.3 Limitations and Future Research

While the results are promising, there are some limitations to the current study. First, the dataset is limited to five technology companies, which may not generalize to other sectors or asset classes. Future research could expand the scope by incorporating a broader range of assets, including commodities, bonds, or cryptocurrencies.

Moreover, the study focuses on transaction costs based on historical price movements. Future work could explore incorporating other factors that influence transaction costs, such as liquidity, trading volume, or market depth, into the GNN model to improve prediction accuracy further.

Another avenue for future research is the application of reinforcement learning to dynamically adjust the portfolio based on real-time market conditions and transaction costs. This could allow for continuous portfolio optimization and potentially higher profitability.

## 6. Conclusion

This paper presents a novel approach to portfolio rebalancing by integrating Graph Neural Networks for transaction cost prediction with Dijkstra's algorithm for path optimization. The results demonstrate that this method can effectively reduce transaction costs and improve the overall efficiency of the rebalancing process. The combination of machine learning and classical optimization algorithms provides a powerful tool for portfolio managers, particularly in high-frequency trading environments.

Future research will focus on expanding the asset universe, incorporating additional cost factors, and exploring reinforcement learning techniques to further enhance the portfolio optimization process.